Fossil Biodiversity: Red Noise Plus Signal

We have examined the Fourier power spectrum of extinction, origination, and total biodiversity in the marine fossil record. We find them inconsistent with self-similarity as well as random walk behavior, but instead dominated by low-frequency power. Even in the background of this low-frequency dominance, the 62 My biodiversity cycle is predominant.

A recent analysis[1] of fossil biodiversity found a 99%-significant 62 My periodicity against permuted random walk Monte Carlo simulations, based on a large database of marine genera[2] with improved dating[3]. Earlier studies have claimed self-similarity in extinction[4] or little inconsistency with random walk behavior[5]. We re-examine some of these issues with additional analysis of the newer data.

Fourier analysis of detrended biodiversity, origination, and extinction produces power spectra P strongly declining with frequency, which is inconsistent with random walk behavior. Analysis of autocorrelations shows them approaching zero at lag 20 My[5] but with considerable structure at larger lag, best studied with spectra. Power spectra of extinction based on improved dating are different from earlier work[4] showing far more fluctuation and no power-law behavior. Earlier smooth power-law results may be a sampling artifact[6]. The Hurst Exponent H can characterize processes with "memory". It is defined by $R/\sigma \propto n^H$, where R is the range of the data, σ is the sample standard deviation, and n is the number of steps; H=0.5 for a pure random walk[7]. Although we do not find a single power law over the whole domain of n, we find that a best fit for extinction, expressed as number or fraction of genera, detrended or not, is H~0.7, again inconsistent with pure random walk behavior. Power spectra of origination and total biodiversity approach $P \propto k^{-2}$, marginally consistent with self-similarity and with the use of the Hurst Exponent; this shows H~1, very far from a random walk. Our analysis of origination, extinction, and biodiversity as a function of geological (sub)stage, rather than time also shows behavior inconsistent with a random walk. All exhibit evidence of long-range memory using both spectral and range analysis.

In Figure 1 we show a log-log plot of the Fourier power spectrum of detrended biodiversity as a function of time. We have cut off at the upper end close to the Nyquist frequency corresponding to the typical length of geological stages. The power spectrum is characteristic of "red noise"[7], low frequency dominated. Even so, the 62 My oscillation peak stands out at better than 99% confidence assuming Gaussian statistics with the given power spectral shape. We have used completely different methods of estimating significance than Rohde & Muller[1], but our results support their conclusions, rather than any suggestion of self-similarity or random walk character to fossil biodiversity, origination, or extinction.

Adrian L. Melott* and Bruce S. Lieberman[†‡]
Email: melott@ku.edu
* Department of Physics and Astronomy
† Department of Geology
‡ Department of Ecology and Evolutionary Biology
University of Kansas, Lawrence, Kansas 66045 USA

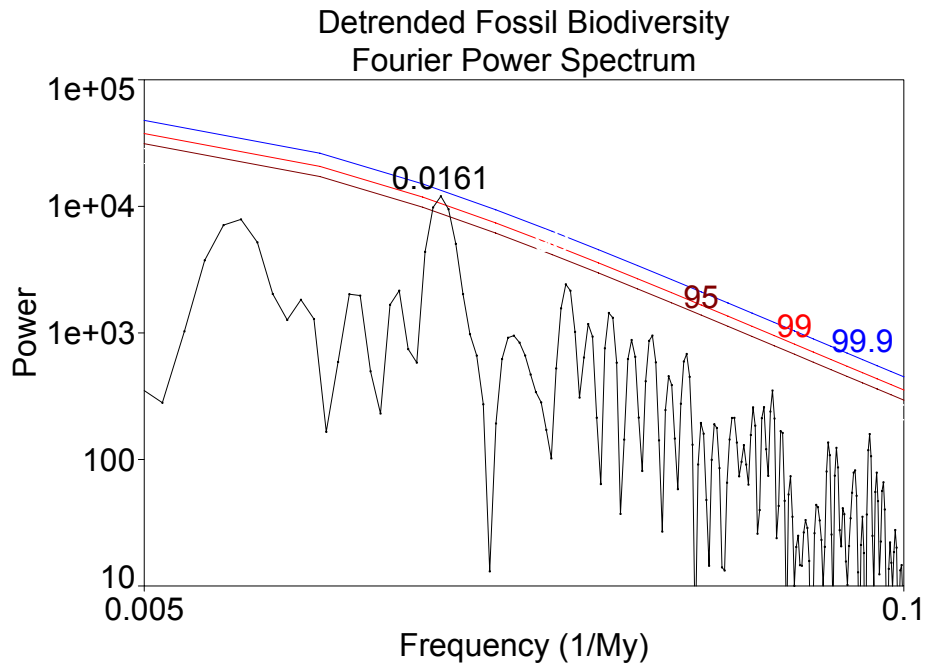

Figure 1: The count of well-resolved marine genera was detrended by a best-fit cubic, and the power spectrum determined as plotted here. The slope over the decade 0.01<f<0.1 is approximately -2. The annotated peak stands out at better than 99% confidence assuming Gaussian statistics for the given spectral shape, and accounts for approximately 35% of the total variance.